\documentclass{article}
\usepackage{spconf,amsmath,graphicx}
\usepackage[table,xcdraw]{xcolor}
\usepackage{booktabs}
\usepackage{graphicx}
\usepackage{amssymb}
\usepackage{hyperref}
\usepackage{multirow}
\usepackage{algorithm}
\usepackage[noend]{algpseudocode}

\title{``It Is Okay To Be Uncommon": Quantizing Sound Event Detection Networks on Hardware Accelerators with Uncommon Sub-Byte Support}

\name{Yushu Wu$^{12}$\sthanks{Work done during internship at Bose Corporation.},
      Xiao Quan$^{1}$, 
      Mohammad Rasool Izadi$^{1}$,
      Chuan-Che~(Jeff) Huang$^{1}$
      }

\address{$^1$ Bose Corporation, MA, USA, 
\\
        $^2$ Northeastern University, MA, USA.
 }

\begin{document}
\abovedisplayskip=0pt
\abovedisplayshortskip=0pt
\belowdisplayskip=0pt
\belowdisplayshortskip=0pt
%
\maketitle

\begin{abstract}

If our noise-canceling headphones can understand our audio environments, they can then inform us of important sound events, tune equalization based on the types of content we listen to, and dynamically adjust noise cancellation parameters based on audio scenes to further reduce distraction. However, running multiple audio understanding models on headphones with a limited energy budget and on-chip memory remains a challenging task.
In this work, we identify a new class of neural network accelerators (e.g., NE16 on GAP9) that allows network weights to be quantized to different common (e.g., 8 bits) and uncommon bit-widths (e.g., 3 bits).
We then applied a differentiable neural architecture search to search over the optimal bit-widths of a network on two different sound event detection tasks with potentially different requirements on quantization and prediction granularity (i.e., classification vs. embeddings for few-shot learning).
We further evaluated our quantized models on actual hardware, showing that we reduce memory usage, inference latency, and energy consumption by an average of 62\%, 46\%, and 61\% respectively compared to 8-bit models while maintaining floating point performance.
Our work sheds light on the benefits of such accelerators on sound event detection tasks when combined with an appropriate search method.


\end{abstract}
\section{Introduction}
\label{sec:intro}


Imagine if our noise-cancelling headphones were like smart companions for our ears. They may intelligently understand the sounds around us to infer acoustic scenes (e.g., on a flight), then dynamically tweak their noise-cancelling parameters to minimize distractions. Furthermore, they may gently inform us whenever something important, like the doorbell, chimes in. They may also optimize the equalization settings based on the specific media content we are experiencing.

Recent progress in the field of deep-learning-based sound event detection (SED), and the broader realm of audio recognition, have brought us closer to realizing this vision. However, due to the size limitations and the expectation for longer streaming time, fitting and running multiple of such models on low-power memory-constrained headphones and earbuds continues to be a challenging task.

One common technique to fit deep neural networks (DNN) on edge devices is through quantization, in particular, 8-bit quantization has been widely used as many micro-controllers can run integer operations efficiently \cite{cmsis-nn}. With quantization-aware training ~\cite{zhou2016dorefa} and knowledge distillation ~\cite{cerutti2019kd}, such an approach reduces memory usage by about 75\% with minimal or no drop in accuracy comparing to floating point originals. Nevertheless, 75\% memory reduction can still be insufficient for more capable but more complex models. To further reduce memory usage and increase energy efficiency, binary-weights-networks (BWN) has also been investigated on SED tasks. While they achieve substantially higher efficiency, they cause the accuracy to drop significantly (e.g., 7\% in ~\cite{cerutti2020sound}), making them unappealing for the aforementioned experiences. 

To identify a more balanced sweet spot between 8-bit and binary quantization, researchers have recently been investigating neural architecture search (NAS) over mixed-precision sub-byte quantization. For example, ARM proposed the UDC framework \cite{fedorov2022udc} to automatically search over network types (e.g., depth and width), bitwidth, and sparsity. Yet, only common power-of-two bitwidths (i.e., 1, 4, 8) were included in the framework as their hardware do not implement and cannot take advantage of uncommon bitwidths operations (e.g., 3 and 5-7 bits). On the other hand, Uhlich et al. \cite{sony-uhlich2019mixed} investigated both common and uncommon bitwidths in their search method. However, they only analyzed the impact on accuracy and model size without considering actual hardware support and overhead. Both works also only evaluated their methods on image recognition tasks, therefore it is still unclear how their results may generalize to SED.

In this paper, we present three contributions. First, we identify a new class of neural network accelerators that supports both common and uncommon sub-byte operations. Second, we apply an efficient NAS to search over the optimal bitwidth per layer of a network and evaluate the impact on actual hardware. We achieve 54-69\% memory reduction, 45-47\% latency reduction, and 53-69\% energy reduction compared to 8-bit models while maintaining floating point performance. Third, we evaluate our method on two different SED tasks –– \textit{generic classification} and \textit{few-shot learning} -- which potentially have different requirements on quantization granularity as the latter network produces embeddings for distance comparison rather than simple classification. To the best of our knowledge, we are the first to evaluate NAS over common and uncommon sub-byte on actual hardware. Our work sheds light on the benefits of such accelerators on SED, allowing designers of audio machine learning systems to make a more informed hardware architecture choice. 

\section{Methods}


\label{sec:methods}

\subsection{New Accelerators with Sub-Byte Support}
Most of the micro-controllers for low-power edge devices to-date do not support uncommon bitwidths and cannot run layers with different bitwidths efficiently due to the overhead to re-scale dynamic ranges between layers. For example, ARM Cortex-M processors only support 8-, 16- and 32-bit integer operations \cite{cmsis-nn}. While newer processors such as Syntiant NDP200 \cite{ndp200} support sub-byte, they are limited to regular power-of-two bitwidths (1, 4, and 8 bits). 

Recently, Paulin and Conti introduced a new set of hardware accelerators including RBE \cite{rbe} \cite{xnor2018} and NE16 \cite{ne16} on the GAP9 processor \cite{gap9}, which open the door for mixed-precision sub-byte quantization on actual hardware. For example, NE16 uses bitwidth in its innermost loop for partial sum computation. While it only allows 8 and 16-bit activation, it supports variable bitwidths for weights. Using its 3x3 Convolution Mode as an example, NE16 can ingest 3 (kernel height) x 3 (kernel width) x 16 (input channels) x 1 bit and 5 (input feature map height) x 5 (input feature map width) x 16 (input channels) x 8 bits to perform 1296 1x8-bit Mutiplication-and-Accumulation in one cycle. In the next cycle it will then handle the next bit in the 3x3x16 weight block, thus for a 3-bit-weight CONV layer, NE16 will simply repeat the computation three times to go over each bit of the weights.

\subsection{Differentiable NAS using FracBit}

To train mixed-precision DNN models, previous works adopt differentiable NAS with DARTS-like approaches (e.g., \cite{fedorov2022udc}) and straight-through estimator (STE) \cite{sony-uhlich2019mixed} to determine the optimal bitwidths. However, DARTS-like search introduces a large search space since it needs to go over all possible bitwidths of a layer, which incurs a long search time. On the other hand, STE avoids such a problem by using a trainable bitwidth parameter. However, since it uses hard bitwidths in forward computation (i.e., rounded a bitwidth parameter to integer) and soft bitwidths in back-propagation (i.e., represents a bitwidth parameter in floating point), this makes it difficult for the gradient to reflect the proper bitwidth direction –– whether to increase or decrease bitwidth. In our experience this makes the training difficult to converge. 

To address the challenges mentioned above, we implemented a recent NAS solution, FracBit~\cite{yang2021fracbits}, which uses fractional bitwidths with interpolation in forward computation and back-propagation. Eq.~\ref{eq:interpolation} illustrates how FracBit work.
In the equation, $n_{\ell}$ represents the bitwidth of a layer $\ell$, $\lfloor \cdot  \rfloor$ and  $\lceil \cdot  \rceil$ are the floor and ceiling function. $f_{n_{\ell}}(x)$ is the mapping function to quantize a value $x$ to ${n_{\ell}}$-bit and back to floating point (with precision loss). 
FracBit explicitly captures the importance of both the higher and lower bitwidths, and the gradient towards each side is easily computed through back-propagation. This strategy increases training time by merely 2$\times$ while allowing consistent convergence. 
\vspace{.3em}
\begin{equation}
    f_{n_{\ell}}(x) \approx (\lceil{n_{\ell}}\rceil-n_{\ell})f_{\lfloor{n_{\ell}}\rfloor}(x) + (n_{\ell}-\lfloor{n_{\ell}}\rfloor)f_{\lceil{n_{\ell}}\rceil}(x)
    \label{eq:interpolation}
    \vspace{.3em}
\end{equation}

To find models that fit tight memory constraints, we use a memory term in the loss function. As shown in Eq.~\ref{eq:size loss}, $w_{\ell\mathtt{n}}$ is the storage footprint of a layer $\ell$ with \texttt{n} bitwidth, $s_{\ell\mathtt{FP}}$ is the space overhead needed to store the scalers for layer $\ell$, and $S_\mathtt{target}$ is the target storage constraint. As shown in Eq.~\ref{eq:total loss}, we then combined the memory loss with an accuracy loss unique for each task to create the final loss function. In this equation, $\beta$ is a hyper-parameter to control their relative importance, and we used $\beta$ = 0.1 in our experiments.
\vspace{.3em}
\begin{equation}
    \mathcal{L}_{\mathtt{size}} = \left|\sum_{\ell}\left(w_{\ell\mathtt{n}}+s_{\ell\mathtt{FP}}\right) - S_\mathtt{target}\right|
    \label{eq:size loss}
    \vspace{.3em}
\end{equation}
\vspace{.0em}
\begin{equation}
    \mathcal{L}_{\mathtt{total}} = \mathcal{L}_{\mathtt{acc}} + \beta \mathcal{L}_{\mathtt{size}}
    \label{eq:total loss}
\end{equation}
\vspace{-2em}

\subsection{Generic \& Few-Shot Sound Event Detection}
Ideally, intelligent headphones need to recognize both \textit{generic} sounds (e.g., voice) as well as \textit{specific} sounds (e.g., a unique doorbell).
The former could be solved with a \textit{generic SED} model that detects pre-defined sound classes, while the latter may be solved by few-shot learning.
These tasks may require different granularity in weights, activation, and output quantization.
For example, prior work on speech enhancement \cite{tinylstm2020} have found that 8-bit mask output is insufficient to achieve good performance and 16 bits is needed instead.
Since few-shot learning produces high-dimensional embeddings for distance comparison instead of the typical logits for softmax, it is unclear how much it can benefit from low bitwidth quantization.
We therefore included both SED tasks in our evaluation.  

In this work, we use a dilated CRNN model (DCRNN) \cite{crnn} for generic SED, as it achieves good performance and is a common baseline. It contains two dilated convolution layers, two bi-directional LSTMs, and a linear layer. For few-shot learning, we use a hierarchical sound event detection and speaker identification network (HiSSNet) \cite{shashaank2023hissnet} as it has been shown that training with few-shot learning and a hierarchical loss improves accuracy. HiSSNet uses MobileNetV2 \cite{sandler2018mobilenetv2} as its encoder. 

\section{Experiments Setup}
\label{sec:exp}

In this section, we will describe our 2$\times$2 experimental setup over two SED tasks, two quantization approaches, and with results on three different performance metrics. 
\vspace{-1em}
\subsection{Datasets}
We replicated the same HiSSNet and dilated CRNN (DCRNN) training and evaluation setup as Shashaank et al. Same as the authors, we aggregated monophonic audio recordings from 7 different datasets (e.g., BBC \cite{bbc}, VCTK \cite{vctk}), resulting in a compiled dataset with 211.6K audio files (673.4 hours). To prepare model inputs, audio files are first down-sampled to 16,000 Hz and down-mixed to single-channel waveform. One-second segments were then randomly sampled from the processed audio files to create a $97\times64$ bin log-Mel spectrogram with a 32ms FFT window and 10ms hop size. The classes of this dataset are organized by a custom three-tier hierarchy with 7-top level classes (e.g., speech), 19 middle-level classes (e.g., female speech), and 1319 lower-level classes (e.g., speaker X). Of the 1319 lower-level classes, 56 classes are for generic SED (e.g., emergency alarm), and the remaining classes capture unique voices of 1263 speakers. We used the whole aggregated dataset for HiSSNet training and evaluation, and removed the speaker subset when training and evaluating DCRNN for generic SED. We used 90\% of the dataset for training and the remaining for validation. 

\vspace{-1em}
\subsection{Training with Sub-byte Quantization}
We implemented two different quantization approaches in training. In addition to FracBit-based DNAS, one naive approach on hardware accelerator that supports sub-byte quantization is to use the same bitwidth to quantize every layer of a network with quantization-aware training (QAT). While prior work such as PACT \cite{choi2018pact} have investigated this approach, they were not perform on SED tasks, nor on real hardware. This serves as a baseline in addition to the floating point models. 

To train DCRNN for generic SED, we followed common mini-batch-based training where each mini-batch can contain any of the 56 low-level SED classes. For few-shot learning tasks, we included all 1319 classes in training and trained using episodic-batches with a 100-episode 12-way 5-shot setup where we sampled 12 classes, 5 query set, and 5 support set in each episode, and repeated the same process 100 times per epoch. Since we had significantly more classes for speakers, we balanced the training by randomly selecting the 12 classes from three different configurations per episodic-batch: SED only, SED \& speaker identification (SID), and SID only. 

We first trained DCRNN and HiSSNet for 100 and 1000 epochs in floating points (the loss had already stalled). For each \textit{fixed-bitwidth} model, we train models with a pre-defined bitwidth for 10 epochs on DCRNN and 100 epochs on HiSSNet. For FracBit, we use DNAS for DCRNN with 5 epochs and HiSSNet with 50 epochs, followed by a fine-tuning step where we rounded and froze the identified bitwidths in the remaining epochs (5 for DCRNN and 50 for HiSSNet).

\vspace{-1em}
\subsection{Metrics}
To evaluate generic SED we randomly sampled 60 samples from the validation set for 100 times and recorded the mean accuracy. For HiSSNet, we used a similar process but instead sampled 120 samples in each round -- 5 support set and 5 query set for each class, for a total of 12 classes -- to obtain the final mean accuracy. Since HiSSNet is designed to support both \textit{generic} and \textit{specific} sound event detection, we looked at two different accuracy metrics: L3 accuracy, which captures performance on \textit{few-shot learning} to detect \textit{specific} sounds, and L1 accuracy that captures performance on \textit{generic SED}.   

\vspace{-1em}
\subsection{Hardware Setup}
We built an evaluation board with a GAP9 processor and several current sense resistors to measure inference latency and energy consumption. The GAP9 processor has a 9-core RISC-V cluster with AI accelerator~(NE16) and a 1-core RISC-V controller.
We first converted each quantized model to C code through nntool~\cite{nntool}, then wrapped it in a test application to feed in test inputs repeatedly. We clocked GAP9 at 370MHz, and used a Saleae Logic Pro 8 \cite{saleae} at 12.5MHz sample rate to log current changes over time. We precisely marked the start and end time of an inference in Logic 2 \cite{logic2} to obtain latency, then derived energy consumption per inference assuming a fixed voltage measured before inference.  

\vspace{-1em}
\section{Results}

\label{sec:results}

\begin{figure}[t]
\vspace{-1em}
    \centering
    \includegraphics[width=\linewidth]{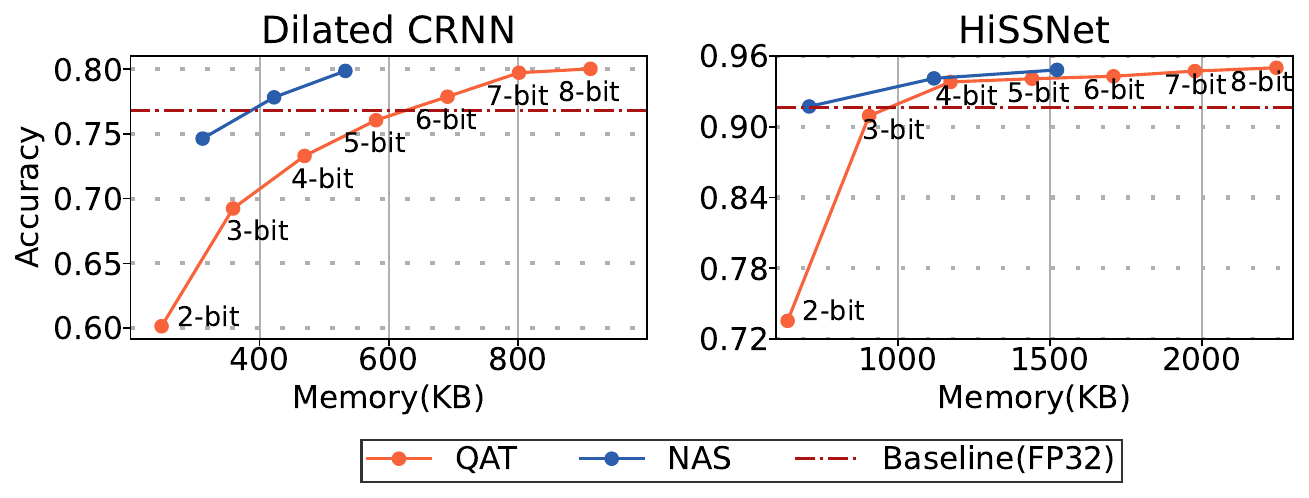}
    \vspace{-2em}
    \caption{
    FracBit~(optimal bitwidth per layer) achieves the best Pareto-front when comparing with fixed-bitwidths models.
    }
    \vspace{-1em}
    \label{fig:memory vs. accuracy}
\end{figure}

\subsection{Fixed-bitwidth achieves decent memory reduction}
We first show that with a naive fixed-bitwidth quantization approach~(wXaY, indicates X-bit weights and Y-bit activations), we achieve a decent balance between memory and accuracy on SED tasks. For example, on generic SED with a DCRNN, we could use 6-bit weights for every layer to reduce memory usage by 24\% compared to the 8-bit model (912KB $\rightarrow$ 691KB) while still maintaining floating point accuracy. Similarly, for few-shot learning with HiSSNet, the model with 4-bit weights reduces memory footprint by 48\% (2246KB $\rightarrow$ 1173B) with no accuracy degradation compared to the floating point model. As prior work showed, the quantized models achieve better performance than floating point especially with high bitwidth (e.g., \cite{choi2018pact}).


\vspace{-1em}
\subsection{FracBit achieves about 20\% more memory reduction}
\vspace{-.5em}
To further improve the balance, we applied FracBit to automatically identify the optimal bitwidth for each layer under different memory constraints. We found the models identified via FracBit achieve the best Pareto-frontiers on memory usage and accuracy -- approximately 20\% more memory reduction compared to the best fixed-bitwidth models, as Fig.~\ref{fig:memory vs. accuracy}. For instance, by setting $S_{\mathtt{target}}=400$ (KB) for DCRNN, we maintain floating point accuracy on generic SED while reducing 54\% memory usage compared to the 8-bit original (912KB $\rightarrow$ 422KB). This represents an additional 30\% (269KB) and 18\% (158KB) memory reduction compared to its 6-bit (no accuracy drop) and 5-bit (0.7\% accuracy drop) counterparts respectively. On HiSSNet, when setting $S_{\mathtt{target}}=700$ (KB), we saw 69\% memory reduction with no accuracy loss on few-shot learning, which is 21\% (465KB) more than its 4-bit variant. This nearly 20\% additional memory reduction is significant as many low-power embedded devices only have about 32-2048KB on-chip flash \cite{cerutti2019kd}.

\begin{table}[t]
\vspace{-.5em}
\centering
\resizebox{0.9\columnwidth}{!}{%

\begin{tabular}{c|c|cccc}
\toprule
Config & Memory(KB)                     & Acc.(\%)$\uparrow$                      & Latency(ms)$\downarrow$                  & Energy($\mu J$)$\downarrow$                     \\
\midrule
                        \cellcolor[HTML]{EFEFEF}FP32               & \cellcolor[HTML]{EFEFEF}3550.0 & \cellcolor[HTML]{EFEFEF}76.83 & \cellcolor[HTML]{EFEFEF}---  & \cellcolor[HTML]{EFEFEF}---    \\ \hline
                        w8a8 & 912.3 & 80.03 & 44.83 & 2707.6 \\
                        w7a8 & 801.6 & 79.73 & 39.14 & 2380.0 \\
                        w6a8 & 690.9 & 77.88 & 35.50 & 2072.0 \\
                        w5a8 & 580.2 & 76.06 & 27.94 & 1714.5 \\
                        w4a8 & 469.5 & 73.30 & 23.92 & 1470.7 \\
                        w3a8 & 347.7 & 69.24 & 19.57 & 1186.5 \\
                        w2a8 & 248.2 & 60.14 & 9.84  & 623.7 \\ \hline
                        \cellcolor[HTML]{C0C0C0}$S_\mathtt{target}=500$ & \cellcolor[HTML]{C0C0C0}532.8  & \cellcolor[HTML]{C0C0C0}79.87 & \cellcolor[HTML]{C0C0C0}27.35 & \cellcolor[HTML]{C0C0C0}1664.8 \\
                        \cellcolor[HTML]{C0C0C0}$S_\mathtt{target}=400$ & \cellcolor[HTML]{C0C0C0}422.4  & \cellcolor[HTML]{C0C0C0}77.83 & \cellcolor[HTML]{C0C0C0}20.73 & \cellcolor[HTML]{C0C0C0}1281.5 \\
\cellcolor[HTML]{C0C0C0}$S_\mathtt{target}=300$ & \cellcolor[HTML]{C0C0C0}311.7  & \cellcolor[HTML]{C0C0C0}74.64 & \cellcolor[HTML]{C0C0C0}16.00 & \cellcolor[HTML]{C0C0C0}968.5  \\
\bottomrule

\end{tabular}%
}
\caption{Dilated CRNN Results. We show that FracBit-based DNAS with various target storage~($S_{\mathtt{target}}$) achieves significant memory, latency and energy reduction compared to fixed-bitwidth~(e.g., w6a8) while still maintaining floating point accuracy.}
\label{tab:crnn_results}
\vspace{-1em}
\end{table}

\begin{table}[t]
\centering
\resizebox{0.9\columnwidth}{!}{%
\begin{tabular}{c|c|ccccc}
\toprule
\multicolumn{1}{c|}{} &                                & \multicolumn{2}{c}{Acc.(\%)$\uparrow$}                            &                               & \multicolumn{1}{c}{}                             \\
\multicolumn{1}{c|}{\multirow{-2}{*}{Config}} & \multirow{-2}{*}{Memory(KB)}   & \multicolumn{1}{c}{L1} & \multicolumn{1}{c}{L3}                             & \multirow{-2}{*}{Latency(ms)$\downarrow$} & {\multirow{-2}{*}{Energy($\mu J$)$\downarrow$}} \\
\midrule
                            \cellcolor[HTML]{EFEFEF}FP32                 & \cellcolor[HTML]{EFEFEF}8551.5 & \cellcolor[HTML]{EFEFEF}93.45 & \cellcolor[HTML]{EFEFEF}91.62 & \cellcolor[HTML]{EFEFEF}---  & \multicolumn{1}{c}{\cellcolor[HTML]{EFEFEF}---} \\ \hline
                            w8a8 & 2245.5 & 97.83 & 95.00 & 11.19 & 499.0 \\
                            w7a8 & 1977.4 & 96.97 & 94.72 & 10.53 & 476.4 \\
                            w6a8 & 1709.4 & 96.53 & 94.28 & 8.94 & 421.1 \\
                            w5a8 & 1441.4 & 96.53 & 94.07 & 8.32 & 397.8 \\
                            w4a8 & 1173.3 & 95.82 & 93.80 & 7.09 & 353.5 \\
                            w3a8 & 905.3  & 92.98 & 90.92 & 5.79 & 309.1 \\
                            w2a8 & 637.2  & 78.92 & 73.55 & 5.50 & 281.4 \\ \hline
                            \cellcolor[HTML]{C0C0C0}$S_\mathtt{target}=1500$ & \cellcolor[HTML]{C0C0C0}1523.6 & \cellcolor[HTML]{C0C0C0}97.67  & \cellcolor[HTML]{C0C0C0}94.82  & \cellcolor[HTML]{C0C0C0}8.69 & \cellcolor[HTML]{C0C0C0}421.9                   \\
                            \cellcolor[HTML]{C0C0C0}$S_\mathtt{target}=1100$ & \cellcolor[HTML]{C0C0C0}1118.7 & \cellcolor[HTML]{C0C0C0}96.66  & \cellcolor[HTML]{C0C0C0}94.10  & \cellcolor[HTML]{C0C0C0}6.12 & \cellcolor[HTML]{C0C0C0}330.2                   \\
\cellcolor[HTML]{C0C0C0}$S_\mathtt{target}=700$  & \cellcolor[HTML]{C0C0C0}708.2  & \cellcolor[HTML]{C0C0C0}93.44  & \cellcolor[HTML]{C0C0C0}91.72  & \cellcolor[HTML]{C0C0C0}5.96 & \cellcolor[HTML]{C0C0C0}317.4                  
 \\ 
\bottomrule

\end{tabular}%
}
\caption{HiSSNet Results. The results applied to few-shot learning with FracBit, with various target storage~($S_{\mathtt{target}}$) achieving significant memory, latency and energy reduction over fixed-bitwidth~(e.g., w3a8) while keeping floating point accuracy.}
\vspace{-1.5em}
\label{tab:proto_results}
\end{table}

\subsection{FracBit achieves 45-47\% latency and 53-69\% energy reduction on NE16}
Being able to quantize networks with uncommon bitwidth is one thing, reaping their benefits on hardware is another. Without hardware that supports sub-byte data representation and operations, we would still need to pad the sub-byte values to byte values, resulting in a loss of efficiency. In this section, we further show that with new accelerators that support sub-byte, we could indeed run these models on-device and achieve significant latency and energy reduction compared to the 8-bit models with no drop in accuracy comparing with floating point models (see Table ~\ref{tab:crnn_results} and ~\ref{tab:proto_results}). For example, we found that the DCRNN trained with FracBit ($S_{\mathtt{target}}=400$) achieves 45\% reduction in inference latency and 53\% in energy consumption on GAP9 with NE16. Additionally, for few-shot learning with HiSSNet ($S_{\mathtt{target}}=700$), we decrease inference latency by 47\% and energy consumption by 69\% compared to the 8-bit model. 

\begin{figure}[t]
    \centering
    \vspace{-.5em}
    \includegraphics[width=0.9\linewidth]{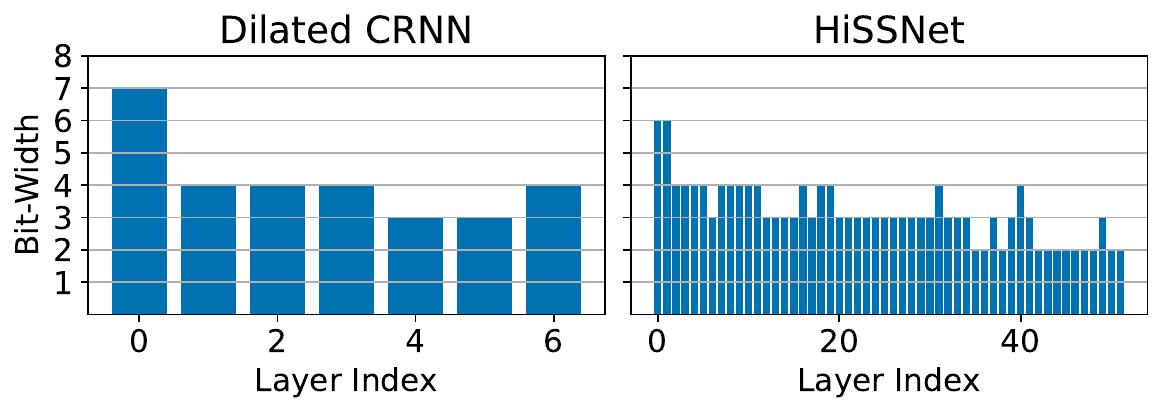}
    \vspace{-1em}
    \caption{The optimal bitwidth per layer for DCRNN and HiSSNet while preserving floating point accuracy.}
    \vspace{-1em}
    \label{fig:optimal bitwidth}
\end{figure}
\vspace{-1em}
\subsection{Uncommon bitwidths are used in searched models}
Finally, we show that the models identified by FracBit did take advantage of the various common and uncommon bitwidths available for their different layers (see Fig.~\ref{fig:optimal bitwidth}). For DCRNN ($S_{\mathtt{target}}=400$) and HiSSNet ($S_{\mathtt{target}}=700$), they both leverage higher uncommon bitwidth (e.g., 7 and 6) for the first layer, following by lower common and uncommon bitwidths (e.g., 3) in the latter layers. 

\vspace{-1em}
\section{Conclusion}
\vspace{-1em}
Embedding deep-learning-based audio understanding and sound event detection (SED) capabilities on noise-cancellation headphones enables many meaningful applications. However, running multiple of such algorithms on headphones with strict power and memory constraints is a difficult task. In this paper, we identify a new class of hardware accelerators that provides common and uncommon sub-byte support. By combining this hardware with an efficient neural network search method (FracBit), we show that we achieve significant memory (54-69\%), inference latency (45-47\%), and energy consumption (53-69\%) reduction on two different sound event detection tasks on real hardware compared to 8-bit models while maintaining floating point performance. Our results provide key insights to designers of audio machine learning systems to consider what neural network accelerators to use in the future.

\bibliographystyle{IEEEbib}
\bibliography{strings,refs}

\end{document}